\documentclass[12pt,doublespacing]{elsarticle}

\usepackage{algorithm}
\usepackage{algorithmic}
\usepackage{amsmath}
\usepackage{amssymb}
\usepackage{array}
\usepackage{balance}
\usepackage{booktabs}
\usepackage{dsfont}
\usepackage{fancyhdr}
\usepackage{graphicx}
\PassOptionsToPackage{hyphens}{url}\usepackage{hyperref}
\usepackage{multirow}
\usepackage{setspace}
\usepackage{subfigure}
\usepackage{tabularx}
\usepackage{times}
\usepackage{url}

\newcolumntype{V}{>{$\vcenter\bgroup\hbox\bgroup}c<{\egroup\egroup$}}

\begin{document}

\begin{frontmatter}

\title{COVID-19 and Science Communication: The Recording and Reporting of Disease Mortality}

\author[STAOA]{Ognjen Arandjelovi\'c}

\address[STAOA]{
  University of St Andrews\\
  St Andrews KY16 9SX\\
  United Kingdom\\
  Tel: +44(0)1334 46 28 24\\
  E-mail: \texttt{ognjen.arandjelovic@gmail.com}\\
  Web: \texttt{http://oa7.host.cs.st-andrews.ac.uk/}\\~\\~\\~\\~\\~\\
  
  {\flushleft
  \normalfont{ The author declares no conflict of interest.\\~\\
    
  The author is the sole contributor to all aspects of the present work, including research, analysis, and manuscript preparation.\\~\\

  There is no funding source to declare.~\\~\\
  
  No ethical approval was needed.~\\
  }}
  \clearpage

}

\begin{abstract}
  The ongoing COVID-19 pandemic has brought science to the fore of the public discourse and considering the complexity of the issues involved, with it also the challenge of effective and informative science communication. A particularly contentious topic, in that it is both highly emotional in and of itself, as well as in that it sits at the nexus of the decision-making process regarding the handling of the pandemic, which has effected lockdowns, social behaviour measures, business closures, and others, concerns the recording and the reporting of the disease mortality. To clarify a point which has caused much controversy and anger in the public debate, the first part of the present article discusses the very fundamentals underlying the issue of causative attribution with regards to mortality, lays out the foundations of the statistical means of mortality estimation, and concretizes these by analysing the recording and reporting practices adopted in England, and their widespread misrepresentations. The second part of the article is empirical in nature. I present data and an analysis of how COVID-19 mortality has been reported in the mainstream media in the UK and the USA, including a comparative analysis both across the two countries as well as across different media outlets. The findings clearly demonstrate a uniform and worrying lack of understanding of the relevant technical subject matter by the media in both countries. Of particular interest is the finding
  that with a remarkable regularity ($\rho>0.998$) the greater the number of articles a media outlet published on COVID-19 mortality, the greater the \emph{proportion} of its articles misrepresented the disease mortality figures.
  \\
\end{abstract}


\end{frontmatter}

\clearpage

\section{Introduction}
The crucial role that science plays in our everyday lives is hardly something that needs to be emphasised even to the general public~\cite{walsh1982public,claessen2005public}. Moreover, in the economically developed world, science is for the most part seen as a positive actor~\cite{macoubrie2004public}: it is science that has helped us avoid or overcome previously widespread diseases and illnesses; it is science that has facilitated our being able to communicate across vast distances using video and audio; it is science that has made travel fast, efficient, and accessible, allowing many to explore relatively cheaply distant parts of the globe; it is science that has made access to large swaths of knowledge freely and readily accessible to most; and so on. Unsurprisingly, polling consistently shows that scientists too are seen in a positive light~\cite{ipsos2017ipsos}. The negative aspects of science, which the public does recognize (correctly or incorrectly), are largely well confined to specific realms: the pace of lifestyle change~\cite{walsh1982public}, applications seen as `playing God' or `playing with nature' (e.g.\ genetic modification, creation of new life forms)~\cite{bredahl2001determinants,allum2008science}, or rogue actors' misuse thereof~\cite{claessen2005public}.

However, over the last two years, that is since the emergence of COVID-19, the place that science plays in our lives appears to have changed substantially. From the largely benevolent supporting actor working in the background, supporting, facilitating, and enhancing various everyday pursuits we undertake, science has come to the fore and is being used to justify -- for better or worse, I state this in a value free sense -- in modern times virtually, if not literally unprecedented restrictions on people's freedoms in countries with historically liberal values. Science is used to justify the prohibition to leave one's residence~\cite{brown2020passport}, to legislate compulsory for the cessation of normal business operations~\cite{gaglione2020covid}, to impose bans on socializing with others~\cite{brown2020passport}, etc. When science is placed at the crux of decision-making that effects such severe harmful effects (I am referring to the aforementioned restrictions themselves only, which are undoubtedly harmful, rather than the net effect thereof which may very well be beneficial), it is unsurprising that the public starts to take interest in the relevant science, and seeks to understand and scrutinize it~\cite{brzezinski2020belief}. Yet, this endeavour is fraught with difficulty. Firstly, considering the breadth and the depth of competence required to understand the relevant processes and phenomena to an extent whereby this understanding (and thereby I am not referring merely to the knowledge of procedural or factual matters, that is veritism~\cite{goldman2015reliabilism}, but actual \emph{understanding}~\cite{pritchard2014knowledge}) is sufficient to facilitate a meaningful critical assessment of experts' views, the notion that this competence could be attained in a short period of time by the general public is rather absurdly na\"ive. Secondly, and perhaps this is where the greatest danger lies, it is this lack of understanding of science and the scientific method, that makes the lay public unaware of the limitations of its knowledge, making ill-founded arguments appear convincing and credible, and decreasing trust in the rational scientific authority~\cite{arandjelovic2021ai,Aran2017c}. In this context, effective science communication is crucial, whether it comes from politicians, scientists themselves, or the media. The overarching message of the present paper is that science communication in these delicate and febrile times have been found wanting. Inept and misleading communication, often stemming from a lack of understanding of the subject matter itself, has resulted in undue (in the specific context considered herein, which is not to dismiss other, possibly correct criticisms~\cite{cooper2021good,lavazza2020role}) public scepticism towards scientific advice, a reluctance to adopt and follow guidance, increased discontent~\cite{gerbaudo2020pandemic}, etc. 

Herein I focus on the specific issue of communication regarding COVID-19 mortality, a particularly emotive topic at the nexus of the decision-making processes which led to the great number of the aforementioned restrictive and far-reaching measures aimed at dealing with the pandemic, and pivotal in shaping the public's attitude and behaviour. I start by discussing the very fundamentals underlying the issue of causative attribution with regards to mortality, that is what it means that a person `has died of' something, in Section~\ref{s:mortality}. Having clarified this notion which has caused much controversy, to say nothing of anger on all sides of the debate played out in public, and in particular having explained why the phrasing is epistemologically inappropriate when applied on a personal level, I lay out the foundations to the statistical means of mortality estimation on the cohort, or population, level in Section~\ref{ss:survival}. In Section~\ref{ss:recordingAndReporting} I concretize these statistical approaches by analysing the COVID-19 recording and reporting practices adopted in England, and explain why certain measures were adopted and how they were misrepresented first and foremost by the media, but also by some scientists when communicating with the public. The second part of the article, namely Section~\ref{s:media}, is empirical in nature. Specifically, I present data and analyse how COVID-19 mortality has been reported in the mainstream media in the UK and the USA, including a comparative analysis both across the two countries as well as across different media outlets. The findings of the analysis clearly demonstrate a uniformly and worrying lack of understanding of the relevant technical subject matter by the media in both countries. Finally, Section~\ref{s:summary} presents a summary of the key points of the article and its conclusions.


\section{On the quantification of mortality rates}\label{s:mortality}
It is unsurprising that the mortality of the novel coronavirus SARS-CoV-2 -- that is the disease it causes, COVID-19 -- has quickly become the primary measure of the virus's direct impact. Mortality is a simple measure in that it is underlain by a binary outcome (death or survival) and it is easily understood by the general public. It is also a highly emotive one.

To a non-specialist, the mortality of a disease also appears as being easily measurable: it is a simple process involving little more than the counting of deaths deemed to have resulted from an infection with SARS-CoV-2. However, underneath this seemingly straightforward task lie a number of nuances. The primary one of these is presented by the question of when a death can be attributed to the virus. Even to laypersons who may not be able to express the reasons behind their reckoning, it is readily evident that this question is different than, say, that of asking when somebody has died of a gunshot wound, e.g.\ as used for the reporting of firearm murders. The key difference between the two stems from the former being a distal and the latter a proximal factor. In seeking a link between distal causes and the corresponding outcomes of interest, the analysis of causality is complicated by the complexity effected by numerous intervening and confounding factors. Thus, to give a simple example, while it may be a relatively straightforward matter to establish respiratory failure as the proximal cause of death, it is far less clear when COVID-19 can be linked to it as the distal cause, and to what extent, especially if the patient has pre-existing conditions. 

The key insight stemming from the above is that indeed, it is fundamentally impossible to claim with certainty that any \emph{particular} death was caused by COVID-19. Rather, the approach has to be on \emph{cohort} (or population) based analysis. The basic idea is reasonably simple: in order to assess how a particular factor of interest affects survival, a comparison is made between cohorts which differ in the aforementioned factor but which are otherwise statistically matched in the potentially relevant characteristics. Indeed, the entire field of study usually termed `survival analysis'~\cite{leung1997censoring}, widely used in biomedical sciences and engineering amongst others, is focused on the development of techniques that can be utilized for such analysis and which are suitable to different scenarios (for example, for problem settings when not all data is observable, or for different types of factor of interest such as discrete or continuous, etc.). In the specific case that we are considering here, the situation is rather straightforward in principle, in that the factor of interest is also binary, namely an individual in a cohort either has had or has not had a positive SARS-CoV-2 diagnosis in the past.

While an in-depth overview of survival analysis is out of scope of the present article, for completeness and clarity it is useful to illustrate just some of the more common methods used to this end in the literature and practice. An understanding of these will help set the ground for the topics discussed thereafter, namely how SARS-CoV-2 deaths should be recorded and how they should be reported, and why the two are different.

\subsection{Survival analysis}\label{ss:survival}

\subsubsection{Kaplan–Meier estimation}\label{sss:KM}
The frequently used Kaplan–Meier estimator~\cite{dimitriou2018principled,gavriel2021assessment,nearchou2021comparison} is a non-parametric estimator of the \emph{survival function} within a cohort. The survival function $s(t)$ captures the probability of an individual's death being no earlier than $t$: 
\begin{align}
    s(t) = Prob(\tau > t),
\end{align}
where $\tau$ is the time of death of an individual in the cohort, treated as an outcome of a random variable. In the simplest setting, the challenge is thus of estimating $s(t)$ given the set $\left\{\tau_j\right\}_{j=1\ldots n}$ where $\tau_j$ is the time of death of the $j$-th individual in a cohort numbering $n$. It is assumed that the outcomes corresponding to different individuals are independent from one another, and identically distributed. However, usually not all $\tau_j$ are available because the time of analysis precedes the death of at least some individuals in a cohort. In our specific example, many of the individuals whose data is analysed, whether previously infected with SARS-CoV-2, will not die for many years in future. To account for this, rather than assuming that all $\tau_j$ are known, the estimate is sought from a set of pairs $\left\{(\hat{\tau}_j, c_j)\right\}_{j=1\ldots n}$ where $c_j$ are the censoring times. The censoring time $c_j$ is the latest time for which the survival or non-survival of the $j$-th individual is known; thus, $\hat{\tau}_j$ is meaningful only if $\hat{\tau}_j \leq c_j$ (e.g.\ the time of death of those individuals who have not died by the time of analysis is not known). It is important to emphasise the assumption that censoring is non-informative, i.e.\ that censoring statistics in both cohorts are identical.

\begin{figure}
  \centering
  \includegraphics[width=0.8\textwidth]{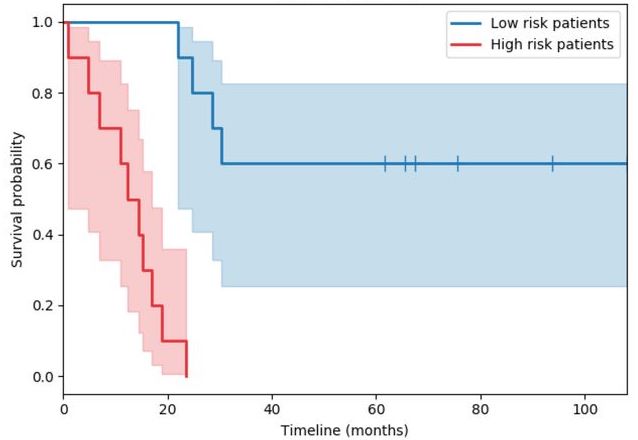}
  \caption{Example of the Kaplan-Meier derived estimate of two survival functions (red and blue lines). The corresponding shared areas indicate the standard deviations of the estimates across time.}
  \label{f:KM_curve}
\end{figure}

It is then a straightforward matter to derive the following estimate for $s(t)$ (for further technical detail and a step by step derivation see e.g.\ the work of Goel \textit{et al}.~\cite{goel2010understanding}):
\begin{align}
    \hat{s}(t) = \prod_{i; t_i<t} \left(1 - \frac{d_i}{n_i} \right)
\end{align}
with $t_i$ a time when at least one death occurred, $d_i$ the number of deaths that happened at time $t_i$, and $n_i$ the individuals known to have survived up to time $t_i$. The plot in Figure~\ref{f:KM_curve} shows a typical example of two survival functions obtained in this manner (the specific example is from a study of immunological features in muscle-invasive bladder cancer). Both estimates, the red and the blue one, start at 1, as all participants in the study are initially alive (and hence, by design, the probability of being alive is 1). Thereafter, the faster decline of the red curve as compared with the blue one, captures a more rapid death rate in the cohort corresponding to the former, i.e.\ a lower probability of survival past a certain point in the future.

\subsubsection{Cox's regression}\label{sss:Cox}
Another widely used technique for survival analysis is Cox's regression (also often referred to as Cox's proportional hazards model), which adopts a somewhat different, semi-parametric approach from that of the Kaplan-Meier estimator in several important ways. Firstly, unlike in the case of the latter, no explicit stratification of the entire patient cohort is performed. Rather, the same effect is achieved statistically. Secondly, the approach is multivariate rather than univariate in nature, which makes the method more appropriate for many real-world analyses when it is not possible to perform randomization or to ensure the satisfaction of other criteria required for Kaplan-Meier analysis.

Cox's regression for survival analysis in its general form is a method for investigating the effect of several variables on the time of death. Central to it is the concept of the \emph{hazard function}, $h(t)$, also called the hazard rate, which is the instantaneous death rate in a cohort (often, it is incorrectly described as the probability of death at a certain time~\cite{wallin2000estimates,malatack1987choosing}). It is modelled as a product of the baseline hazard function, $h_0(t)$ and the exponential of a linear combination covariates (that is, factors of interest), $\left\{ x_i \right\}_{i=1\ldots p}$:
\begin{align}
    h(t) &= h_o(t) \times \exp{ \left\{ w_1 x_1 + w_2 x_2 + \ldots w_p x_p \right\} }\\
    &= h_o(t) \times \exp{ \sum_{i=1\ldots p} w_i x_i }
\end{align}
The coefficients $\left\{ w_i \right\}_{i=1\ldots p}$ can be interpreted as quantifying the effect of the corresponding covariates and can be inferred from data by means of partial maximum likelihood estimation over all observed deaths. The resulting \emph{hazard ratios} $\left\{ \exp{w_i} \right\}_{i=1\ldots p}$ provide a simple way of interpreting the findings, with values around 1 indicating a lack of effect of the factor, and those greater or lesser than 1 respectively, increased and decreased associated hazard of death. A factor in this analysis can be, for example, the presence of a specific disease (e.g.\ COVID-19), a particular demographic characteristic, a comorbidity of interest, etc. Thus, Cox's analysis allows us to interrogate the data as regards the effect of specifically, say, a past positive test for COVID-19, on mortality, adjusted for other factors which too may affect it.

Note that both methods described, namely both Cox’s regression and Kaplan–Meier estimation, are statistical and therefore phenomenological in nature, as opposed to mechanistic -- neither approach models the underlying processes that effect the connection of observable predictor data (e.g.\ the presence of a historical positive test for COVID-19, or one's sex, age, etc.) with the outcome of interest (time of death).

\subsection{The recording and reporting of deaths}\label{ss:recordingAndReporting}
An understanding of the technical basics of survival analysis covered in the previous section erects a solid basis for the consideration of how the recording and the reporting of deaths due to a distal cause should be performed. Indeed, the latter have attracted considerable attention and criticism.

The recording of SARS-CoV-2 deaths during the ongoing pandemic has varied across different jurisdictions in a multitude of ways. My focus here is not on the many practical aspects of this process (e.g.\ how deaths in different settings such as hospitals, homes, and care homes are aggregated, etc.), as important as these are, but on its fundamental, methodological underpinnings which are unaffected by geographical, social, and similar factors. In this regard, we find rather more uniformity so I will use England as a representative example.


\subsubsection{England}\label{sss:england}
Up to August 2020, for England the COVID-19 Data Dashboard reported all deaths in people who have had a prior laboratory confirmed positive SARS-CoV-2 test. Thereafter, two further indicators were included, namely the numbers of deaths of individuals ($\alpha$) who have had their first positive test within 28 days of dying, and ($\beta$) either who have had their first positive test within 60 days of dying \emph{or} who have had COVID-19 on the death certificate (as recorded by a registered medical practitioner). 

The reasoning underlying these choices and the evolution of the reporting system is straightforward to understand. The original intention was to ensure a high degree of confidence in there having been a SARS-CoV-2 infection in a deceased person; hence the requirement of a laboratory confirmed positive SARS-CoV-2 test. The subsequent expansion of the reporting parameters can be seen as an attempt to account for those deceased individuals, possibly many of them, who have been infected but were never actually tested for the virus. It is difficult to argue that these reporting choices are anything other but sensible. It is in the use and the interpretation of them that the nuance lies, as I discuss shortly. 


\subsubsection{Survival analysis...again}\label{sss:survivalAgain}
Although as I noted earlier, the techniques outlined in Section~\ref{ss:survival} provide a good basis for contextualizing the recording of SARS-CoV-2 related deaths in England, it is important to observe that neither Kaplan-Meier estimation nor Cox's regression can be applied in the context just described out of the proverbial can -- further thought is needed to adapt the methods to the problem at hand. In particular, note that as described both approaches are prospective in nature, in that observation begins at a set time for a known cohort. In contrast, in the problem setting of interest the question is retrospectively posed. This challenge is not new and can be addressed in a principled and robust manner by extending the original statistical models. The relevant technical details are involved and not necessary to go into detail herein -- the interested reader would be well advised to consult the work of  Prentice and Breslow~\cite{prentice1978retrospective} or Copas et al.~\cite{copas2001incorporating} for example. 

Nevertheless, there is an important a difficulty of a practical nature which emerges in the context of COVID-19 and which is often one that one has to contend with in many other epidemiological settings. Specifically, the selection of non-diseased individuals is far from straightforward. The reason for this lies in the potentially asymptomatic presentation of the disease. As a consequence, there is a possibility of the non-diseased cohort actually containing individuals who have had COVID-19 at some point, but were unaware of it. An important yet subtle observation that is key to make here is that this data contamination does not act merely so as to reduce the accuracy of analysis or reduce the uncertainty of the conclusions. Rather, there is a \emph{systematic} bias which is introduced. In particular, observe that because it is the asymptomatic individuals, i.e.\ those with the least disease severity and thus the most optimistic prognosis, who are removed from the COVID-19 positive cohort, the overall prognosis of the nominal COVID-19 positive cohort is made to appear worse than it would have been had the asymptomatic cases been included. As a corollary, any analysis applied is likely to produce an overestimate, rather than an underestimate or an unbiased estimate, of COVID-19 mortality. Including a model of the source of bias in the overall statistical model is difficult because the key variables underlying the phenomenon are latent by their very nature.

\subsubsection{Piecing it all together}
My closing remark in Section~\ref{sss:england}, namely that the COVID-19 reporting choices in England are eminently sensible may have resulted in some readers raising their brow. Indeed, when the reporting protocol was first published, the sciolist mainstream (as well as non-mainstream) media was quick to point out that somebody who died after being hit by a bus and who has had a recent positive COVID-19 test, would be included in the reported numbers. This has been widely repeated and a few examples serve well to illustrate the gist of the argument. Thus, the Daily Mail, the highest circulation daily newspaper in the UK following the Sun, reported~\cite{DailyMail}:
\begin{quote}
``...if, for example, somebody tested positive in April but recovered and was then hit by a bus in July, they would still be counted as a Covid-19 victim.''
\end{quote}
Rowland Manthorpe, an editor at Wired magazine who has written for the Guardian, Observer, Sunday Telegraph, Spectator, etc., speaking for Sky News~\cite{SkyNews} echoed the thoughts:
\begin{quote}
``Essentially, there is no way to recover, statistically. So, if I tested positive for COVID-19 today and then I got hit by a bus tomorrow, then COVID-19 would be listed as my cause of death.''
\end{quote}
The reiteration was not limited to media personalities. For example, in an article provocatively entitled ``Are official figures overstating England's Covid-19 death toll?'' the Guardian~\cite{Guardian} reported that an unnamed Department of Health and Social Care source summed up the process as: 
\begin{quote}
  ``You could have been tested positive in February, have no symptoms, then be hit by a bus in July and you’d be recorded as a Covid death.”
\end{quote}

At first sight, these criticisms do not seem entirely unreasonable. Why \emph{would} a person who was killed by being hit by a bus be counted as a COVID-19 death? The Guardian, to its credit, did report an apparent defence by a `source at Public Health England', who was quoted as saying that:
\begin{quote}
  ``...such a scenario would `technically' be counted as a coronavirus death, `though the numbers where that situation would apply are likely to be very small'.''
\end{quote}
This is a rather feeble response; and a misleading one too. The implication is that those deaths indeed \emph{should not} have been recorded but that their infrequency renders the matter of little practical significance. That is incorrect. If these deaths are indeed entirely confounding, statistical analyses such as those outlined earlier, would have found them to be such -- in these cases, they would indeed present as noise in the data and be practically insignificant. However, there is another possibility, which is particularly important when dealing with novel and poorly understood diseases. Imagine if statistical analysis \emph{did} reveal that previously COVID-19 positive people die in greater numbers by being hit by vehicles than their disease free counterparts; in other words, that there was statistical significance to this observation. This would have suggested possibly new knowledge about the virus and its effects. For example, it could have indicated that the virus has long-lasting neurological effects which would affect one's ability to respond in traffic. Indeed, now we \emph{do} know that SARS-CoV-2 is a neurotropic virus with a whole host of neuropathological effects including dizziness, decreased alertness, headaches, seizures, nausea, cognitive impairment, encephalopathy, encephalitis, meningitis, anosmia, etc.~\cite{montalvan2020neurological}. Yet, an explanation of this kind was woefully missing from the mainstream coverage. Also, still working within the premises of this hypothetical scenario, even before any mechanistic understanding is developed, the mere new knowledge that in some way or another, past COVID-19 positivity predicts pedestrian deaths in traffic, as any knowledge, can only be advantageous. For example, it could lead to timely advice to the public to take additional care in appropriate situations. 

To summarize, the recording of COVID-19 related deaths should indeed include all deaths, whatever their proximal cause may be or appear to be, of individuals tested positive any time in the past and these numbers should be reported to the relevant bodies. Thus, if anything, the criteria for the recording of COVID-19 related deaths were insufficiently rather than excessively inclusive. However, the \emph{reporting}, i.e.\ the communication of COVID-19 mortality to the general public should based on robust statistical analyses on the cohort level. It should be a cause of profound concern that even such publications as Scientific American failed in observing what is little more than rudimentary rigour as regards this~\cite{SciAm}, incorrectly claiming:
\begin{quote}
``Nearly 800,000 people are known to have died {\bf of COVID-19}. [all emphasis added]''
\end{quote}

\section{Analysis of mainstream media }\label{s:media}
Having equipped ourselves with the understanding of how data on disease effected mortality ought to be collected and recorded, and interpreted and reported, we are now in good position to turn our attention to the analysis of the reporting of COVID-19 mortality by the mainstream media. I do so in the present section. Its structure follows the usual pattern -- I start by describing my data collection protocol in Section~\ref{ss:data}, then report my analysis of this data in Section~\ref{ss:analysis}, and finally discuss the findings in Section~\ref{ss:discussion}.

\subsection{Data collection}\label{ss:data}
Data was collected from the web sites of the top  circulation national daily newspapers in the UK and the USA, as well as the web sites of the two leading UK television broadcasters, namely the BBC (British Broadcasting Corporation) and ITV (Independent Television; legally Channel~3). Sister publications, e.g.\ the Sun and the Sun on Sunday or the Daily Mail and Mail On Sunday, were considered jointly under the name of the main brand.
The collection of all data was conducted on the 23rd of August 2020. In particular, I recorded the number of articles on each web site containing the exact phrases `deaths with COVID', `deaths from COVID', `deaths of COVID' (n.b.\ the search was not case sensitive). A summary of the raw data can be found in Tables~\ref{t:dataUK} and~\ref{t:dataUSA}.

\begin{table}[]
  \centering
  \def\arraystretch{1.5}%
  \begin{tabular}{c|l||r|rrr}
    \toprule[2pt]
  	 & Brand name & `with' & `from' & `of' & `from'+`of'\\
  	 \hline
     \multirow{6}{*}{\rotatebox{90}{Newspapers}} 
     &The Metro     &   96 &   431 &  2820 &  3251\\
     &The Sun	    &   59 &   416 &   288 &   704\\
     &Daily Mail    &  463 &  9120 & 16000 & 25120\\
     &The Times     &   36 &   147 &   275 &   422\\
     &The Mirror	& 3360 & 10700 & 41400 & 52100\\
     &The Telegraph &  222 &   627 &   788 &  1415\\
     \hline
     \multirow{2}{*}{\rotatebox{90}{TV}} &
     BBC	        &   4210 & 5850 & 4200 & 10050\\
     &ITV	        &   1650 & 5860 & 3440 &  9300\\
     \bottomrule[2pt]
  \end{tabular}
  \caption{Summary of UK data. }
  \label{t:dataUK}
\end{table}

\begin{table}[]
  \centering
  \def\arraystretch{1.5}%
  \begin{tabular}{c|l||r|rrr}
    \toprule[2pt]
  	 & Brand name & `with' & `from' & `of' & `from'+`of'\\
  	 \hline
     \multirow{10}{*}{\rotatebox{90}{Newspapers}} 
     &USA Today        &  64 &  657 &   861 &  1518\\
     &WSJ              &   1 &  250 &  3780 &  4030\\
     &NY Times         &   4 & 5250 & 16600 & 21850\\
     &New York Post    &   1 &  403 &   612 &  1015\\
     &LA Times         &   1 & 1540 &  9720 & 11260\\
     &Washington Post  &  91 & 2670 &  2000 &  4670\\
     &Star Tribune     & 246 & 2380 &  2480 &  4860\\
     &Newsday          &   0 &  263 &  1140 &  1403\\
     &Chicago Tribune  &   9 & 2400 &  7780 & 10180\\
     &The Boston Globe &   7 &  229 &   872 &  1101\\
     \bottomrule[2pt]
  \end{tabular}
  \caption{Summary of USA data. }
  \label{t:dataUSA}
\end{table}

\subsection{Data analysis}\label{ss:analysis}
I started by looking at the proportion of articles by the publication which used the incorrect `deaths of COVID' or `deaths from COVID' phrasing. For the UK based publishers, averaged across publishers, that is, looking at:
\begin{align}
    \frac{1}{n_{publishers}} \sum_{i=1}^{n_{publishers}} \frac{n_i^{incorrect}}{n_i^{correct}+n_i^{incorrect}},
\end{align}
where $n_{publishers}$ is the number of different publishers, and $n_i^{correct}$ and $n_i^{incorrect}$ the number of publisher $i$'s articles with the correct (`with COVID') incorrect (`from COVID' or `of COVID') phrasings, 89.4\% of articles contained an incorrect formulation, with the corresponding standard deviation being 9.0\%. In the USA, the corresponding average was found to be 98.8\%, with the standard deviation of 1.8\%. Taking into account the different numbers of articles which contained any of the search terms, that is:
\begin{align}
    \frac{ \sum_{i=1}^{n_{publishers}}  n_i^{incorrect}} {\sum_{i=1}^{n_{publishers}}  \left( n_i^{correct}+ n_i^{incorrect} \right)},
\end{align}
the overall proportion of incorrectly phrased articles in the UK was 91.0\% and in the USA 99.3\%.

Using the thee one-sample Kolmogorov-Smirnov test~\cite{kaner1980critical}, the distributions of per-publisher proportions of incorrectly phrased articles were confirmed to be log-normal, both for the UK and the USA, at the confidence levels $p=0.0197$ and $p=0.0078$ respectively. The two-sample Kolmogorov-Smirnov test~\cite{friedrich1998computation} further confirmed that the UK and the USA distributions different significantly, that is the null hypothesis of the two sample sets coming from the same log-normal distribution was rejected at the confidence level $p=0.0058$.

\begin{figure}[]
 \vspace{-20pt}
    \centering
    \subfigure[]{\includegraphics{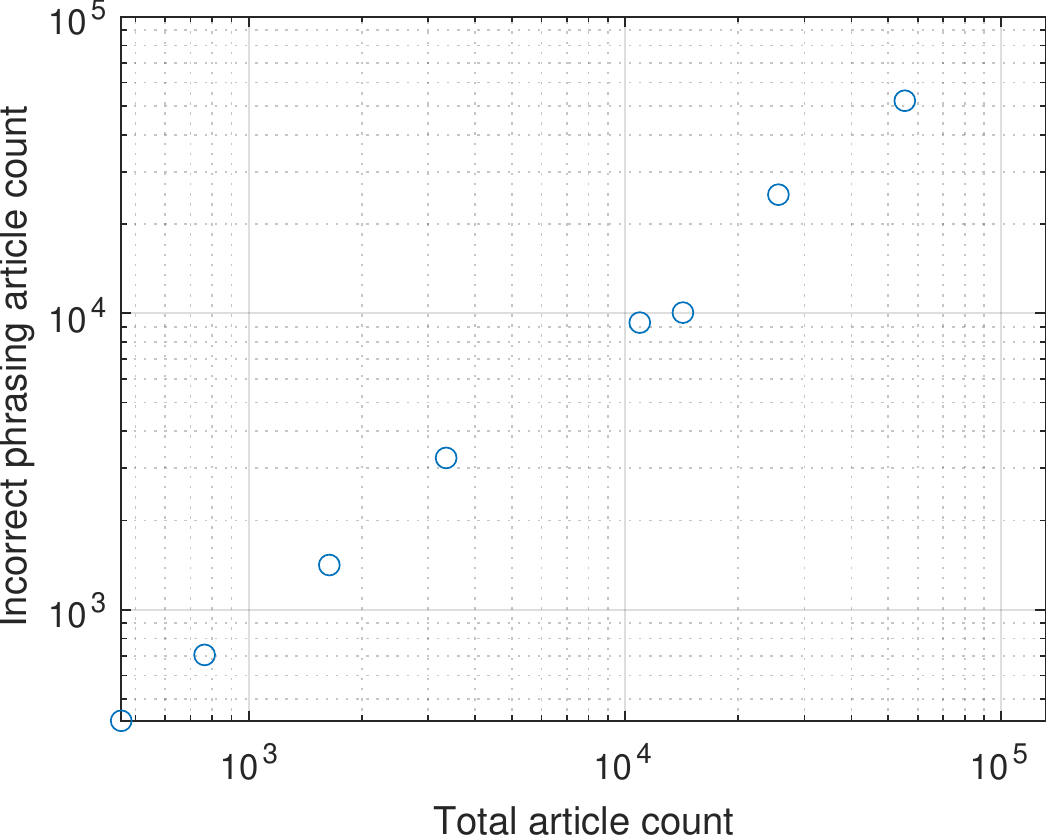}\label{f:ukPlot}}
    \subfigure[]{\includegraphics{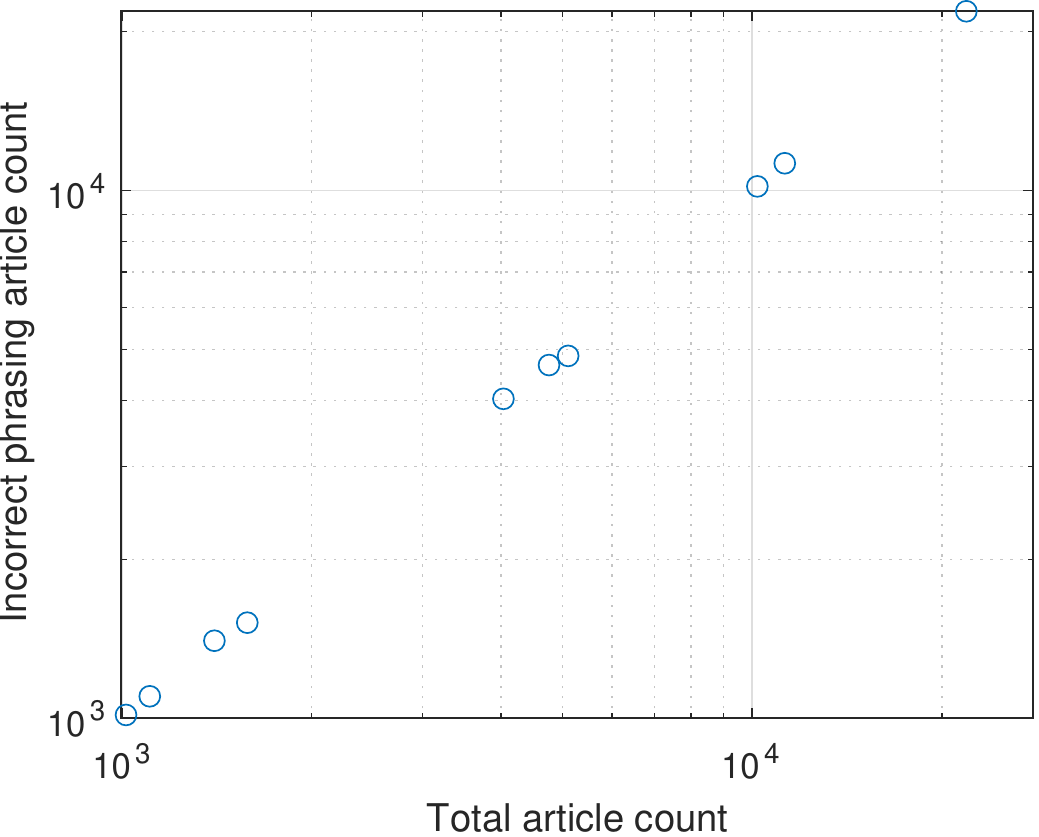}\label{f:usaPlot}}
    \caption{Relationship between the number of incorrectly phrased articles and the total number of articles reporting COVID related deaths for the (a) UK and (b) the USA. A highly linear behaviour is readily observed, with the corresponding Pearson's $\rho$ equal to $0.998$ and $1.000$ respectively. }
\end{figure}

Lastly, the plots in Figure~\ref{f:ukPlot} and Figure~\ref{f:usaPlot} show the relationships between the number of incorrectly phrased articles and the total number of articles reporting COVID related deaths for the UK and the USA respectively. The corresponding Pearson's $\rho$ was found to be $0.998$ for the UK and $1.000$ for the USA.

\subsection{Discussion}\label{ss:discussion}
To begin with a few brief but important comments on the data collection, firstly it should be noted that for the same of uniformity and like-for-like comparison, all data was collected on the same day, namely the 23rd of August 2020, and it includes all historical articles, i.e.\ all articles published on or before that date. Although the exact date in question was selected at random, it was deliberately chosen not to be too early in the pandemic, as a reasonable argument could have been made that journalists, largely unequipped with the kind of expertize needed to understand the communicate the highly technical and to everyday reporting novel kind of information, needed a period of adjustment and learning. This possible objection is fully addressed by the choice of the 23rd of August 2020, which is some 9 months following the first identification of COVID-19 in December 2019~\cite{huang2020clinical} and nearly 6 months following the declaration of a pandemic by The World Health Organization (WHO) on the 11th of March 2020~\cite{WHO}.

The most immediately apparent finding of my analysis in the previous section is that of the strikingly high proportion of articles which incorrectly described COVID-19 deaths as being `from' or `of' COVID-19. Perhaps even more remarkably, this was found to be the case across all analysed media outlets and both in the UK and the USA. Interestingly though, although in both countries all but a few articles used the incorrect phrasings, the transgression in communication was significantly worse in the USA (as I also confirmed statistically in the previous section) -- in the UK approximately 1 in 11 articles did use the correct `with' phrasing, whereas in the USA it was fewer than 1 in 100. Indeed, USA's Newsday did not have a single correctly worded article despite reporting on COVID-19 mortality in 1403 articles, and WSJ, New York Post, and LA Times had only a single correctly worded article each, despite respectively 4030, 21850, and 11260 articles on the topic. The only significant outlier, in a positive sense yet with a still exceedingly high proportion of incorrectly phrased articles of approximately 70.5\%, is the BBC; the BBC was approximately twice as likely to use the correct phrasing in reporting COVID-19 mortality than the next best outlet and more than 3 times as likely as the analysed UK media outlets on average. None of the USA outlets stand out from the rest, even the most accurate one, namely USA Today, using the correct wording in only approximately 4\% of its articles. 

Perhaps the most concerning finding of my analysis concerns the relationship between the proportion of incorrect reporting by a media outlet and the outlet's volume of COVID-19 mortality reporting, summarized by the plots in Figure~\ref{f:ukPlot} and Figure~\ref{f:usaPlot}. As stated in the previous section, we find that with a remarkable regularity (Pearson's $\rho$ for the UK and the USA outlets being $0.998$ and $1.000$ respectively) the greater the number of articles an outlet published on COVID-19 mortality, the greater the \emph{proportion} (n.b.\ not the absolute number, which would be expected) of articles which used an incorrect phrasing with respect to it. There are different reasons which could explain this. For example, it is possible that outlets whose journalists' collective values are more caution driven have as a consequence of that published more on the topic. This would make the transgression -- for a transgression has certainly taken place in the sense that an incorrect claim was made -- entirely of an intellectual, rather than ethical nature. On the other hand, it is impossible to dismiss the possibility of a more sinister cause, such as increased fear oriented reporting being driven by commercial or other interests. Numerous other, more complex reasons are possible too. Considering that the available data do not allow us to favour one hypothesis over another, it would be inappropriate to speculate on the topic; nevertheless, it is important to note the trend and highlight it as an important one for future research. Lastly, note that my analysis found no relationship between the circulation of a newspaper and the corresponding COVID-19 mortality reporting phrasing accuracy.

\section{Summary and conclusions}\label{s:summary}
The coronavirus disease 2019 pandemic has placed the science in the spotlight. However, science does not exist as an immaterial Platonic form -- research is performed by humans, science based decisions are made by humans, and science is communicated by humans. Human fallibility cannot be taken out of science. Scientists, decision-makers, and communicated make not only genuine errors, but also have egos, compete for jobs and prestige, exhibit biases, hold political and broader philosophical beliefs, etc., all of which affect how science is materialized and used. A failure to recognize faults in these processes when they occur can only serve to undermine the general public trust in science and its application~\cite{nielsen2020most,cooper2021good}.

In this article I focused on a specific and highly relevant aspect of science communication in the context of the COVID-19 pandemic, namely that of the disease mortality reporting. Considering the amount of confusion and anger across the spectrum that the issue has caused, I began with a discussion of what it means to say that a person has died \emph{from} a disease. In particular, I explained that while this may be an acceptable phrasing in some circumstances, strictly speaking it is not a meaningful one and as such should generally be avoided. Indeed, this nuance is crucial in situations when there is a significant interaction between different causal factors, as is the case of COVID-19 mortality. Hence, I clarified how mortality ought to be assessed, namely on a cohort (population) basis, and summarized the technical fundamentals which underpin the necessary analysis used to arrive at such estimates. The developed insight was concretized with an analysis of the controversy causing COVID-19 mortality recording and reporting process in England. I showed how the decisions made were widely misunderstood and incorrectly interpreted in the media, and demystified the decisions underlying the process.

The second part of the article turned its attention to the quantitative analysis of COVID-19 mortality reporting by the mainstream media, namely by the top circulation daily newspapers in the UK and the USA, as well as the UK's two leading TV broadcasters, the BBC and the ITV. The results are striking: in both countries the vast majority of articles incorrectly reported COVID-19 deaths, erroneously attributing a causative link between COVID-19 and the death of any person with a past positive COVID-19 test. This effect was observed with a remarkable uniformity over the different outlets considered, with the transgression was significantly worse in the USA than the UK (approximately 89\% vs 99\% respectively).

The effects of poor science communication of the kind considered in this paper must not be underestimated. Not only does incorrect information provide a faulty basis for individual decision-making, but can also penetrate the highest levels of legislative and executive branches of government. Indeed, I leave the reader with three poignant examples. The first of these is from the address to the nation in February 2021 by the president of the USA, Joe Biden~\cite{Biden}: 
\begin{quote}
  ``Today we mark a truly grim, heartbreaking milestone -- 500,071 dead.'',
\end{quote}
the second one from the UK's Prime Minister Boris Johnson statement on coronavirus delivered on the 26th of January 2021~\cite{BJ}:
\begin{quote}
  ``I am sorry to have to tell you that today the number of deaths recorded {\bf from Covid} in the UK has surpassed 100,000... [all emphasis added]'',
\end{quote}
and the last one from the speech by Keir Starmer, then Leader of the Labour Party, at the Labour Party Conference 2021~\cite{Starmer}:
\begin{quote}
  ``We have now lost 133,000 people {\bf to Covid}. [all emphasis added]''
\end{quote}

\bibliography{./refs}
\end{document}